\documentclass[runningheads]{svmult} 
\usepackage{makeidx}  
\usepackage{graphicx} 
\usepackage{subeqnar} 
\usepackage{multicol} 
\usepackage{physprbb} 
\makeindex            
\def\eqcite#1{(\ref{#1})}  
\newcommand{\beq}{\begin{equation}}  \newcommand{\eeq}{\end{equation}}
\newcommand{\barr}{\begin{eqnarray}} \newcommand{\earr}{\end{eqnarray}}
\def\little{\tiny }
\def\tdoor{\tau_{\hbox{\little Door}}} \def\tjump{\tau_{\hbox{\little J}} }
\def\tlife{\tau_{\hbox{\little L}} } \def\ttunnel{\tau_{\hbox{\little T}} }
\def\tzeno{\tau_{\hbox{\little Z}}} \def\tpass{\tau_{\hbox{\little P}}}
\def\tresp{\tau_{\hbox{\little R}}} \def\tpulse{\tau_{\hbox{\little PM}}}
\def\remark#1{\begingroup  \addtolength{\baselineskip}{-.1\baselineskip}
  \addtolength{\lineskip}{-.9\lineskip} \noindent \sf Remark:~ #1 \endgroup}
 \newcommand{\draftlabel}[1]{ }    
\def\delt{\delta \mskip -1mu t }  \def\negonemu{\mskip -1mu}
\def\dddot#1{\vbox{\ialign{##\crcr
          \hbox{~.$\negonemu$.$\negonemu$.}
          \crcr\noalign{\kern 1.1pt\nointerlineskip}
          $\hfil\displaystyle{#1}\hfil$\crcr}}}
\def\Re{\mathop{\rm Re}\nolimits} \def\Im{\mathop{\rm Im}\nolimits}
\def\coltwovector#1#2{\left({#1\atop#2}\right)}

\def\ltsim{\hbox{\kern.25em\raise.5ex\hbox{$<$}\kern-.75em\lower.5ex     \hbox{$\sim$}\kern.25em}}
\def\compC{\hbox{\bf C}} 
\def\tTh{\widetilde\Theta} \def\tW{\widetilde W}
\begin{document}
\title*{Jump time and passage time: the duration of a quantum transition}
\toctitle{Jump time and passage time: the duration \protect\newline of a quantum transition}
\titlerunning{Jump time and passage time}
\author{L. S. Schulman}
\authorrunning{L. S. Schulman}
\institute{Physics Department, Clarkson University, Potsdam, NY 13699-5820, USA}
\maketitle              

\section{Introduction}  \label{sec-introd} \draftlabel{intro}
It is ironic that experimentally time is the most accurately measured physical quantity, while in quantum mechanics one must struggle to provide a definition of so practical a concept as time-of-arrival. Historically, one of the first temporal quantities analyzed in quantum mechanics was {\it lifetime}, a property of an unstable state. The theory of this quantity is satisfactory in two ways. First, with only the smallest of white lies, one predicts exponential decay, and generally this is what one sees. Second, at the quantitative level, one finds good agreement with a simply derived formula, the Fermi-Dirac Golden rule, \draftlabel{golden}
\begin{equation}
\Gamma=\frac{2\pi}\hbar \rho(E) |\langle f |H|i\rangle|^2 \;. 
\label{golden}
\end{equation}
Eq.\ (\ref{golden}) uses standard notation. $\Gamma$ is the transition rate from an initial (unstable) state $|i\rangle$ to a final state $|f\rangle$. The transition occurs by means of a Hamiltonian $H$. The density of (final) states is $\rho$, evaluated at the (common) energy of the states $|i\rangle$ and $|f\rangle$. In terms of $\Gamma$, the lifetime is $\tlife=1/\Gamma$.

The lifetime $\tlife$ is not a property of any one atom (or whatever), but rather of an \emph{ensemble} of like atoms. For much of the twentieth century this was sufficient. One was taught not to inquire too closely about the time evolution of an individual member of an ensemble. An exception to this informed neglect arose as technology allowed experimentalists to focus on transitions in individual atoms \cite{plenio}. Although one can recast these phenomena in ensemble terms, the ensemble is typically conditioned on the fact of the ultimate decay of the system studied. But a similar extension of naive ensemble interpretations was already present in studies of \emph{tunneling time}. The barrier penetration phenomenon of quantum mechanics was sufficiently provocative in its denial of classical notions that one sought places where conventional ideas \emph{could} be applied, for example, trying to assign a time of passage through the barrier. This subject has a long history and a collection of recent views can be found in Ref.~\cite{implications}. Again, in principle, for barrier penetration one deals with ensembles, but if one measures passage time there would need to be conditioning on the fact of the transition, observations of individual transits and a time interval measured for each. Our notation for tunneling time (without distinguishing among the many definitions) is~$\ttunnel$.

The tunneling time concept allowed further probes of the Copenhagen view of quantum mechanics. A decaying particle, for example a nucleus in the Gamow model of alpha decay, was said to undergo a \emph{quantum jump}. The idea (I guess) was that you could measure the particle in its initial state or in its final state. But getting from one to the other was a ``jump." It took a measurement to distinguish one state from the other, putting the jump itself beyond the scope of quantum mechanics, or at least of ordinary unitary time evolution. However, if one could ascribe to the particle a trajectory under the barrier, along with a time during which the particles tunnels, then one has made the first steps in the analysis of this ``jump." Assigning a trajectory is problematic \cite{yamada}, although several authors have used the Feynman path integral \cite{absence,sokolovski,fertig}, acknowledging the limitation that the path contributions only add as amplitudes, not probabilities.

A different ``quantum jump" was exhibited experimentally in the eighties \cite{exp1,exp2,exp3}. This involved an atomic transition for a \emph{single} atom. There wasn't any ``path under a barrier" and as indicated the notion of ensemble needed updating. In particular, one could no longer muddle the distinction between an abstract ensemble and the large number of atoms participating in decay experiments.

In these experiments \cite{exp1,exp2,exp3} one monitored the atom closely, noting when it was in its excited state. The duration of its stay in the excited state was (quantumly) random, and repetition of the experiment gave statistics that could be used to evaluate the lifetime. But from the data it was evident that something else was happening---the famous or infamous jump---and that its time scale, if any could be defined for it, had little to do with lifetime.

The question that I raised \cite{quick} was whether one could say anything about the time interval that elapses between finding the atom in one state and finding it in the other. One does not need the drama of \cite{exp1,exp2,exp3} to ask this question. Radium ($^{226}$Ra) has a half-life of about 1600$\,$yr and one can imagine putting a single such nucleus in an inert matrix and waiting to see an $\alpha$-decay. (This is similar to the experiments cited, where seeing \emph{nothing} meant that the system was still in its metastable state). The interval between being in one state and being in the other is certainly brief. But is it, as early and perhaps loose interpretations of measurement theory would have it, instantaneous? Another example goes back to arguments for the quantized nature of light, as demonstrated by the photoelectric effect. The ``instantaneous" appearance of electrons when ultraviolet light was turned on was a blow to classical interpretations \cite{eisberg}. But again, ``instantaneous" is a matter of technology, and the bounds on this time interval were only about a nanosecond.

In this article I will define two times, each related to the question asked. But they are in general quantitatively different from one another. The times are called \emph{jump time} and \emph{passage time}. Roughly speaking, the first measures how long it takes for the transition process to get seriously underway and the second how long it takes to complete the process in a single exemplar. But it is better not to use too much verbal description. From the definitions below and from the applications, the relevance of each should emerge.

The jump time is designated by the symbol $\tjump$. In the next section I will motivate my definition and arrive at a quantitative expression. The considerations parallel arguments arising in the quantum Zeno effect (QZE). The formula for $\tjump$ turns out to be the next simplest thing you could construct from the Hamiltonian after \eqcite{golden}.

The passage time, designated $\tpass$, has a precise mathematical definition, although in a specific experimental situation it will depend on the apparatus as well as on the system undergoing the transition. It arises from a bound on the minimum time for a state to evolve (with given Hamiltonian) to a state orthogonal to itself.

Definitions are tested by what you can do with them, what they unify. I will show that $\tjump$ arises in several contexts. It is a generalization of tunneling time ($\ttunnel$). It satisfies a kind of time-energy uncertainty relation. For certain transitions it establishes the experimentally observed time scale, although for atomic decays it is immeasurably short. Passage time is related to a theoretical bound found by Fleming \cite{fleming}, a bound not hampered by requiring notions of what a ``measurement" is supposed to be. Although with respect to my own ideas on quantum measurement theory $\tpass$ may prove more significant than $\tjump$, its dependence on the measuring apparatus limits its general applicability.

Finally, because of the many characteristic temporal quantities that will be defined in this article, I have included Table~\ref{Table1} for reference.

\begin{table}
\caption{ Characteristic times}
\begin{center}
\renewcommand{\arraystretch}{1.4}
\setlength\tabcolsep{5pt}
\begin{tabular}{llll}
\hline\noalign{\smallskip}%
Time & Name & Description  \\
\noalign{\smallskip}
\hline
\noalign{\smallskip}

$\tlife  $ & Lifetime       & Usual lifetime for decay, $\hbar/2\pi \rho(E) |\langle f
                              |H|i\rangle|^2$ \\
$\tzeno  $ & Zeno time      & Inverse of energy spread,
                              $\hbar/\sqrt{\langle\psi| (H-E_\psi)^2 | \psi\rangle}$\\
$\tjump  $ & Jump time      & $\tzeno^2/\tlife$ \\
$\ttunnel$ & Tunneling time & As in barrier penetration \\
$\tpass  $ & Passage time   & Minimum time to go from a state to a $\perp$ one, $\pi\tzeno/2$ 
                              \\
$\tresp  $ & Response time  & A property of monitoring apparatus \\
$\tpulse $ & Pulse time     & Interval between ideal pulsed measurements (cf.\ QZE) \\
$\tdoor  $ & Door time      & Metaphorical \\

\hline
\end{tabular}
\end{center}
\label{Table1}
\end{table}

\section{Jump Time} \label{sec:jumptime}    \draftlabel{jumptime}

How long does it take to walk through a doorway? Call this time $\tdoor$. Consider the following experiment. A stream of people passes through a door, one at a time. From time to time, and without looking, I fire a marshmallow across the doorway. Anyone hit by a marshmallow must turn back. Assume the marshmallow crosses the doorway instantaneously. If I fire $N$ times during a time interval of duration $T$, then I expect to turn back a fraction $N\tdoor/T$ of the people. An experiment to measure $\tdoor$ would consist of gradually increasing the marshmallow firing rate until no one can cross. The estimate for $\tdoor$ would then be $T/N$. In other words, when my firing rate reaches $1/\tdoor$ I stop the traffic. Without further refinements this measurement would not define $\tdoor$ by better than a factor two, that is, it defines a {\it time scale}, rather than a precise time.

The same perspective motivates the definition of quantum jump time. The decay, or other quantum transition, corresponds to getting through the door. The process-terminating interruption is an ``observation," a quantum measurement. As for tunneling time, the use of classical concepts means that the doorway analogy is incomplete.

We formalize the discussion: at intervals $\delt$ project onto the initial states, i.e., measure whether the system is still in its initial state. If these disturbances do not slow the decay, then $\delt$ is to be considered longer than the jump. On the other hand, if these projections do slow the decay, then they have reached its time scale. In this way I arrive at a context similar to that of the quantum Zeno effect (``QZE") \cite{zenoreview}.

Let the system begin in a state $\psi$ and let the full Hamiltonian be $H$. After a time $\delt$, $\psi$ evolves to $\exp (-iH\delt/\hbar) |\psi\rangle$. One checks for decay by applying $\langle\psi|$. The probability that it is still in $\psi$ is $ p(\delt) = |\langle\psi| \exp (-iH\delt/\hbar) |\psi\rangle|^2$. A short calculation shows that \draftlabel{expansionone}
\beq
p(\delt)= 1- \left(\frac{\delt}{\tzeno}\right)^2 +\hbox{~O}(\delt^4)
\label{expansionone}
\eeq
where \draftlabel{zenodef}
\beq
\tzeno^2 \equiv
   \frac{\hbar^2}{\langle\psi| (H-E_\psi)^2 | \psi\rangle}  \label{zenodef}
\eeq
and $E_\psi \equiv \langle\psi|H|\psi\rangle$. I call $\tzeno$ the ``{\it Zeno time}," notwithstanding my lack of full concurrence with the classical allusion \cite{dominated}.

\medskip

\remark{It is worth taking a second look at the derivation of \eqcite{expansionone}, since the appearance of high frequency terms in the off-diagonal matrix elements has exercised some authors {\rm\cite{kurizki}}. Let
$$f(t) \equiv \langle\psi| \exp (-i(H-E_\psi)t/\hbar) |\psi\rangle       \;.$$
First, assume that this function has at least three derivatives in $[0,t]$, so that in particular, besides $E_\psi$, $\langle\psi|H^2|\psi\rangle$ and $\langle\psi| H^3 |\psi\rangle$ must be finite. Then by standard theorems, one can write $f(t)=1 -(t/\tzeno)^2/2 +t^3 \dddot f\!\!(t^*)$ for some $t^*$ between 0 and $t$. Calculating $|f|^2$ (to get $p(t)$) shows the deviation from $1-t^2/\tzeno^2$ to be no larger than O$(t^3)$. When a fourth derivative exists, $\Re\!\dddot f(0)=0$ implies \eqcite{expansionone}.}

\medskip

Now suppose that many projections are made during a time $t$, carried out at intervals $\tpulse$. Then to leading order, at $t$, the probability of being in $\psi$ is \draftlabel{interrupt}
\beq
p_{\,\hbox{\little Interrupted}}(t)
=   \Bigl[p(\tpulse)\Bigr]^{t/\tpulse}
 \approx \left[ 1- \left(\frac{\tpulse}t \frac  {t\tpulse}{\tzeno^2}\right)
                     \right]^{ t/\tpulse}
   \approx \exp\left(-t\tpulse/\tzeno^2 \right)      \;.    \label{interrupt}
\eeq
To define a jump time, we want to know whether this differs from standard decay. Without projections the probability for being in $\psi$ is \draftlabel{uninterrupteddecay}
\beq
p_{\,\hbox{\little Uninterrupted}}(t)= \exp(-t/\tlife) \label{uninterrupteddecay}
\eeq
with $\tlife\equiv1/\Gamma$ the usual lifetime (``$\Gamma$" of \eqcite{golden}). Comparing \eqcite{interrupt} and \eqcite{uninterrupteddecay}, we see that the interrupted decay will be slower for $\tpulse < \tzeno^2/\tlife$ \cite{reverse}. We are thus led to define the ``jump time" as the time for which the slowdown would begin to be significant, namely \draftlabel{jumptimedef}
\beq
\tjump  \equiv \tzeno^2/\tlife   \;.  \label{jumptimedef}
\eeq
In words, $\tjump$ is the time such that if one inspected a system's integrity at intervals of this duration, the decay would be slowed significantly \cite{quick,durat}.

\medskip

\remark{Because my goal is only to define a {\it time \underline{scale}}, I do not attempt greater precision. For example, in \eqcite{uninterrupteddecay}, because of the initial quadratic dependence, one may want to change the extrapolated time-zero value. For our purposes, however, the normalization is irrelevant, since it is the decay \emph{rate} whose equality fixes $\tjump$.}

\medskip

\remark{Recall that \eqcite{golden} uses the first moment of the Hamiltonian. The jump-time definition, \eqcite{jumptimedef}, involves the second moment, in a way, the simplest step beyond minimal decay information.}

\section{Corroborations of the definition} \label{sec:corroboration} \draftlabel{corroboration}

The usefulness of jump time will be demonstrated in a number of contexts: 1) comparison with tunneling time; 2) time-energy uncertainty principle; 3) reconciling continuous measurement with the QZE; 4) experiments on the quadratic regime of decay.

\subsection{Comparison with tunneling time} \label{sec:comparison} \draftlabel{comparison}

In \cite{dominated} a simple example of quantum tunneling was studied in an effort to estimate $\tzeno$. There is an interesting complication in this calculation, namely the dependence of $\tzeno$ on the initial state ($\psi$ of \eqcite{zenodef}). This complication is the reflection of a recurrent problem: what is a metastable state? For $\tlife$ this question is not critical, since by the time the exponential decay sets in, transients have disappeared. But now it is the transients we study. Our choice in \cite{dominated} was to minimize the second moment of the Hamiltonian, hence to maximize $\tzeno$. With this approach we found, with fairly rough approximations, that 
\beq  \tzeno^2=\tlife\ttunnel      \;. 
\eeq
Comparing this to \eqcite{jumptimedef}, it is seen that for this kind of transition, the tunneling time is the jump time.

\subsection{Time-energy uncertainty principle} \draftlabel{uncertainty}\label{sec:uncertainty}

An interpretation of $\tjump$ in terms of bandwidth and uncertainty relations can be found by combining \eqcite{zenodef}, for $\tzeno$, with \eqcite{golden}, for lifetime, $\tlife$. After some manipulation one obtains  \draftlabel{bandwidth}
\beq
\tjump  =\frac{\tzeno^2}{\tlife}
        =\frac 1
       { \int \frac{dE}{2\pi\hbar} \frac {\rho(E)}{\rho(E_\psi)}
             \frac
            { |\langle E|H-E_\psi|\psi\rangle|^2}
            {|\langle f|H  |\psi\rangle|^2}}  \;. \label{bandwidth} 
\eeq
Because of the orthogonality of the initial and final states, one can insert a ``$-E_\psi$" into the Golden rule matrix element. Thus the ratio \draftlabel{bandratio}
\beq
\frac{\rho(E)}{\rho(E_\psi)} 
   \frac { |\langle E|H-E_\psi|\psi\rangle|^2}
       {|\langle f|H       |\psi\rangle|^2}       \label{bandratio}
\eeq
is of order unity when $E$ passes through $E_\psi$. As $E$ moves away from $E_\psi$ a variety of patterns is possible, depending on the specific physical situation. One scenario is for this ratio to become smaller, mainly because with increasing energy deviation, $|E\rangle$ becomes rather different from $|i\rangle$ \cite{otherscenario}. In any case, this ratio, whose numerator incorporates transitions to all possible on-shell and off-shell states, measures the ability of the Hamiltonian $H$ to move the system away from its initial state. One thus has a band of accessible transition states.

With this perspective, $\tjump$ is (the inverse of) an integral over energies (or frequencies) of an order unity-function describing the modulation of the lowest band of accessible states. It follows that $\tjump $ is the inverse bandwidth for the transition. This is a completely reasonable conclusion: you would like to create a situation where the system's transition is sudden. Your success is governed by the frequencies available. The accessibility of those frequencies is the essence of the bandwidth. This makes the jump time a reflection of a kind of time-energy uncertainty relation. As such it is a statement of this relation that is consistent with the views expressed in \cite{hilgevoord}.

\subsection{Reconciling continuous measurement with the QZE} \draftlabel{reconciling}\label{sec:reconciling}

The sequence of infinitely rapid projections envisioned in the usual derivation of the Quantum Zeno Effect (QZE) is hardly the way measurements actually take place. Mostly they could be described as ``continuous," in the following sense. An apparatus monitors a system and when some particular event takes place it is triggered and reports that event. Before that report, the apparatus, by its silence, is telling you, "No, the event has not yet taken place." If this picture is true, then one should expect \emph{all} decay to be suppressed, since the unwavering attention of the apparatus should act like a continuous check---effectively with a zero time interval between measurements---that no decay has taken place.

This problem was addressed some years ago by several authors \cite{sudbery,kraus,pereszeno,perescontin}, some of whom also wished to dispense with the (perhaps metaphysical) traditional notion of ``measurement" and instead include the apparatus as part of the quantum system. They found that adding apparatus-like terms to the Hamiltonian could stop or slow the decay.

In recent work \cite{contin}, I found that the important criterion for determining which ``continuous" measurements could affect decay (or any transition) was a comparison of two quantities: the response time of the apparatus and the jump time of the system being measured. The essential physical idea is that no measurement is ``instantaneous" and any apparatus represents a sequence of physical processes, first getting the signal to the apparatus and then having the apparatus register that signal---the latter typically involves irreversible amplification. What I found was that when the response time of the apparatus, $\tresp$, was on the order of $\tjump$ the decay would be hindered. In particular, an apparatus with response time $\tresp$ had the same effect in slowing decay as idealized pulsed measurements with pulse time $\tpulse=4\tresp$. Moreover, from the development of Sec.~\ref{sec:jumptime} of the present article, $\tpulse$ should be less than or equal to $\tjump$ for there to be a significant effect. Consequently the same criterion (ignoring the factor 4) should hold for $\tresp$.

The demonstration proceeds by making a model of a decay plus an apparatus that ``continuously" monitors that decay. The model Hamiltonian and wave function for the decay alone are \draftlabel{defH}
\beq
H=\pmatrix{ 0      & \Phi^\dagger \cr
            \Phi   & \omega       \cr}
\qquad\quad\hbox{and~}\quad
\psi=\coltwovector xy       \;.     \label{defH}
\eeq
where $x\in\bf C$, $\Phi$ and $y$ are complex column vectors of the same dimension, and $\omega$ is a diagonal matrix. The Schr\"odinger equation (with $\hbar=1$) becomes
\draftlabel{schrodeq}
\beq
 i\dot x = \Phi^\dagger y \;,\qquad
 i\dot y = \omega y +  \Phi x    \;.      \label{schrodeq}
\eeq
One can derive the decay rate from Eq.\ \eqcite{schrodeq} by assuming the time de\-pen\-dence $\exp[-i(E-i\Gamma/2)t]$ for both $x$ and $y$. One obtains
\beq
E-i\frac\Gamma2  = \Phi^\dagger {1\over E-\omega -i\Gamma/2}\Phi
\longrightarrow
 \int d\omega {\rho(\omega) |\phi(\omega)|^2 \over
               E-\omega -i\Gamma/2}
\eeq
where the arrow indicates a continuum limit, $\rho$ is the density of states, and $\phi$ the appropriate limit of $\Phi$. The usual manipulations now give $\Gamma=2\pi \rho(0) |\phi(0)|^2$, the Fermi-Dirac Golden Rule. The Zeno time for the state with $x=1$ (and $y=0$) is simply $\tzeno=1/\sqrt{\Phi^\dagger \Phi}$.

The Hamiltonian in \eqcite{defH} can be thought of as describing a two level atom coupled to the electromagnetic field. For $\psi^\dagger=(x^*,0)$ the atom is in the unstable state (call this level \#1), while $\psi^\dagger =(0,y^\dagger)$ describes the decayed atom (in level \#2) with photon(s) emitted. As a monitoring device we imagine another system coupled to the atom that allows the atom to decay once more (to atomic level \#3), emitting one or more additional photons, providing sufficient decoherence for this to be considered a measurement. The coupling strength between levels 2 and 3 will be thought of as adjustable (perhaps some function of an external electric field). Such a model is embodied in the following Hamiltonian
\draftlabel{Hwithapp}
\beq
H=\pmatrix{ 0   & \Phi^\dagger & 0              \cr
          \Phi  & \omega       & \Theta^\dagger \cr
            0   & \Theta       & W              \cr}  \;.  \label{Hwithapp}
\eeq
The additional levels, $\{W\}$, can be thought of as the apparatus and $\Theta$ is the 2-3 coupling. We assume that the levels are numerous enough and so distributed that the transition induced by this coupling is effectively irreversible.

To see how the combined system behaves we make a substitution similar to that done above: all components of the wave function are given the time dependence $\exp(-izt)$. One obtains \draftlabel{eigvalapparatus}
\beq
z= \Phi^\dagger
   \frac1{z-\omega - \Theta^\dagger \frac1{z-W}\Theta}
   \Phi
\;. \label{eigvalapparatus}
\eeq
In the usual way (which was implicit above), $1/(z-W)$ is evaluated using the formula $1/(x\pm i\epsilon) = P(1/x) \mp i\pi\delta(x)$. We write the result as \draftlabel{gammatheta}
\beq
\Theta^\dagger\frac1{z-W}\Theta
   = \Delta E -i\frac{\Gamma_\theta}2  \;. \label{gammatheta}
\eeq
This formula uses the reasonable assumption that $\Theta$ does not depend on which photon was emitted in the 1-2 transition. $\Gamma_\theta$ is the essential descriptor of the apparatus, indicating the rate at which it takes the atom from level \#2 to level \#3. The inverse of $\Gamma_\theta$ is thus the response time of the apparatus, which we denote $\tresp$. Formula \eqcite{gammatheta} is inserted in \eqcite{eigvalapparatus} to yield
\draftlabel{eigvalapparatus2}
\beq
z= \Phi^\dagger \frac1{z-\omega -\Delta E +\frac i{2\tresp}} \Phi
\;. \label{eigvalapparatus2}
\eeq
We next assume that the response time is \emph{so} small that its inverse dominates the $z-\omega-\Delta E$ term in the denominator of \eqcite{eigvalapparatus2}. The imaginary part of $z$ is thus a transition rate \emph{away from the initial excited state, in the presence of the observing apparatus}. Writing $\Im z=-i\Gamma_{\hbox{\tiny effective}}/2$, \eqcite{eigvalapparatus2} implies
\draftlabel{effective}
\beq
\Gamma_{\hbox{\tiny effective}}=\frac {4\tresp}{\tzeno^2} \label{effective}
\eeq
(using $\Phi^\dagger \Phi\, \tzeno^2=1$, which is still true for the full $H$, including the apparatus). If $\Delta E$ is itself comparable to $1/\tresp$ there is a slight modification of \eqcite{effective}, reducing $\Gamma_{\hbox{\tiny effective}}$, but unless $\tresp\Delta E\gg1$ this does not change our qualitative conclusions.

The expression \eqcite{effective} is to be compared to the effective decay rate when under \emph{pulsed} idealized observation, as conventionally described in the \hbox{QZE}. From our Eq.~\eqcite{interrupt}, this rate is $\tpulse/\tzeno^2$. Comparing this with
\eqcite{effective}, we see that \emph{the same degree of hindrance is obtained for an apparatus with response time} $\tresp$ \emph{and pulsed measurements (projections) at intervals} $\tpulse$, \emph{provided}
\beq
\tresp=\tpulse/4  \;.
\eeq
Moreover, as discussed in Sec.~\ref{sec:jumptime}, neither interruption will slow the decay unless it is $\ltsim \tjump$.

\medskip

\remark{Once one deals with Hamiltonians and ordinary unitary evolution (rather than mysterious wave function ``collapses") both for the ``system" and for the ``apparatus," another perspective is opened for understanding the hindering of decay because of continuous, rapid-response, observation \cite{hinder}. One starts with a system (with Hamiltonian \eqcite{defH}), which has a continuum into which to decay. Coupling a detector to this can be thought of as changing the spectral properties of the combined system. In particular what it can do is push the energy of the excited level and the continuum into which it decayed away from one another. Thus the halting of decay occurs because there are no longer levels that match (including the photon energy) the energy of the excited atom. This is discussed in \cite{hinder} and \cite{contin}.
A continuous version of the \emph{anti-}QZE \cite{reverse} has a corresponding explanation.}

\subsection{Experiments on the quadratic regime of decay} \draftlabel{experiment}\label{sec:experiment}

Atomic and nuclear transitions take place quickly, putting the times discussed in this chapter out of reach of contemporary measurement for those systems. In \cite{quick,durat} I estimated that for atomic transitions $\tjump\sim10^{-20}\,$s. However, there is a recent experiment \cite{raizen} where the potential seen by the particles, including a barrier, has a distance scale of a few hundred nm. This experiment, a measurement of Landau-Zener tunneling, has (for us) two benefits: the time scales are \emph{much} longer and the potential can be quickly modified.

The experiment \cite{raizen} consists of putting ultra-cold Na atoms in opposing laser beams that have a relatively small frequency difference between them. As a result the potential seen by each atom is time-dependent. Going into the atom's accelerated frame, the potential can be written $V=V_0 \cos(2k_Lx)+aMx$ (``tilted washboard"), where $a$ is the acceleration arising from the frequency mismatch. Initially a small value of $a$ is given to get rid of atoms not caught in the potential, after which it is sharply increased, giving rise to the tunneling situation. It is then switched off in such a way that it is possible to deduce what fraction of the atoms has escaped from the potential. For long times this quantity dies exponentially with a time scale of 70$\,\mu$s. However, for short times it is demonstrably \emph{not} exponential---it begins with what appears to be zero slope, tilts a bit, and then after roughly 5 to 10$\,\mu$s goes over to the exponential form.

In \cite{lzjump} I showed how one could get a back-of-the-envelope estimate of the duration of this transient period. Recall that my derivation of the jump time, $\tjump$, was essentially a playoff of the quadratic and exponential time dependencies (ignoring finer nuances of the decay curve). Hence it should provide an estimate of the duration of the transient period in the experiment just described.

To make this estimate it was not necessary to calculate either of the quantities $\tzeno$ or $\tlife$. Instead I appealed to the interpretation of $\tjump$ as inverse bandwidth, Eq.~\eqcite{bandwidth}. Which states are accessible to the atom in this potential? In fact it is a periodic potential and the atom is initially in its lowest band. If it were not for the tilt, the states in this band would be eigenstates of the Hamiltonian. The tilt couples these states and makes the otherwise stable states unstable. I thus take the band of accessible states to be just the band of Bloch states. But the width of this band can be calculated from the period of the potential and the mass of the atom. The bandwidth is just \draftlabel{band}
\beq
E_b = \frac{\hbar^2 K^2}{2M} \;, \label{band}
\eeq
where $M\simeq23 M_p$ and the wavenumber is $K = 1/94\;$nm \cite{niu}. We evaluate \draftlabel{expttime}
\beq
\tau_{_J} = \frac\hbar{E_b} 
     = \frac{2M}{\hbar K^2} \simeq 6\;\mu\hbox{s}  \;. \label{expttime}
\eeq
Comparing this to Figures 3 or 4 in \cite{raizen}, it can be seen that the agreement is excellent. In evaluating \eqcite{expttime} there are \emph{many} powers of 10, and I found it remarkable that they condense to \emph{any} reasonable result, much less one that was close to the actual experiment \cite{timescales}.

\medskip

\remark{The closeness of the evaluated time in \eqcite{expttime} to the experimental result should be considered fortuitous. My estimate depends on the wavelength of the light and the mass of the particle. It does \emph{not} explicitly depend on the strength of the potential nor on the rate of acceleration, features that are known to affect the duration of the non-exponential decay.}

\section{Passage time} \label{sec:passagetime}\draftlabel{passagetime} 
\subsection{Fleming's bound and the Ersak equation} \draftlabel{flemingbound}  \label{sec:flemingbound}

Given a Hamiltonian $H$ and a state $\psi$, define $U(t) \equiv \exp[-i(H-E_\psi)t/\hbar]$, with $E_\psi \equiv \langle\psi|H|\psi\rangle$. We define a quantity related to what Fleming \cite{fleming} calls the \emph{integrity} amplitude
\draftlabel{integrity}
\beq
f(t)\equiv \langle\psi|U(t)|\psi\rangle    \;.       \label{integrity}
\eeq
Next, the function $\phi_t$ is defined to be that portion of the evolute that is orthogonal to $\psi$:
\draftlabel{phidef}
\beq
U(t)|\psi\rangle=f(t)|\psi\rangle+|\phi_t\rangle  \;, \label{phidef}
\eeq
with $\langle\psi|\phi_t\rangle=0$. Successive application of $U(t)$ and $U(t')$ to $\psi$, followed by left multiplication by $\psi^\dagger$, leads to
\draftlabel{multiply}
\beq
f(t+t')=f(t)f(t')+\langle\psi|U(t')|\phi_t\rangle   \;. \label{multiply}
\eeq
Using the variable $-t'$, the adjoint of \eqcite{phidef} is
\beq
\langle\psi|U(t')=\langle\psi|f^*(-t')+\phi_{-t'} \;.
\eeq
Multiply this equation on the right by $|\phi_t\rangle$ to yield
$\langle\psi|U(t')|\phi_t\rangle = \langle\phi_{-t'}|\phi_t\rangle$.
When this is substituted in \eqcite{multiply}, we get
\beq
f(t+t')=f(t')f(t)+ \langle\phi_{-t'}|\phi_t\rangle   \;.  \label{ersak}
\eeq
Fleming calls this the Ersak equation. Take the derivative of \eqcite{ersak} with respect to $t'$, set $t'$ to zero, and use the fact (from \eqcite{integrity}) that $\dot f(0)=0$ to yield \draftlabel{fdot}
\beq
\dot f(t)=-\langle \dot\phi_0 | \phi_t\rangle \;.  \label{fdot}
\eeq
From the derivative of \eqcite{phidef} it is clear that
\beq
\langle\dot\phi_0|\dot\phi_0\rangle=\frac1{\hbar^2}\langle\psi|(H-E_\psi)^2|\psi\rangle
         \equiv  \frac{(\Delta H)^2}{\hbar^2}  \;.     
\eeq
For convenience we write $f\equiv g\exp(i\gamma)$, with $g$ real and non-negative and $\gamma$ real. We apply the Schwarz inequality to \eqcite{fdot}:
\beq
|\dot f| \leq \frac{\Delta H}\hbar \sqrt{1-g(t)^2}  \;.
\eeq
Using $|\dot f|^2=\dot g^2+\dot\gamma^2g^2$, we immediately have  
\beq
|\dot g| \leq \frac{\Delta H}\hbar \sqrt{1-g^2}  \;. 
\eeq
Finally, letting $g\equiv \cos\theta$ provides a bound on $\dot\theta$, specifically, $|\dot \theta|\leq \frac{\Delta H}\hbar$. Since $g$ starts at 1, $\theta$ starts at 0, and it follows that
\beq
\theta(t)\leq \frac{\Delta H}\hbar t  \;.
\eeq
This gives our desired bound. Recalling the definition of $g$, it shows that no state can become orthogonal to itself in less than $\pi\hbar/2\Delta H$. But this last quantity is just $\pi\tzeno/2$, in our earlier notation.

This result was derived by Fleming \cite{fleming}, and leads us to define the passage time, $\tpass \equiv \pi\tzeno/2$.

To confirm that the bound can in fact be attained, let $H=\alpha^2\sigma_x$ ($\vec\sigma=$ Pauli spin matrices), and $\psi=|+\rangle$. Then $\tzeno=\hbar/\alpha$, and the system turns over in $\pi\tzeno/2$. This example however does not clarify the relations among the many times that have been defined in this article. Because there is no exponential decay in this case, $\tlife$ is not clearly defined. If one takes it to be the time to first extinction, then $\tzeno$, $\tlife$, $\tpass$ and $\tjump$ are all essentially the same.

The example of the last paragraph is realized in a real-world system: molecules that can exist in two isomers. In practice those molecules for which the transition time between isomers is anything but microscopic appear as one or the other isomer, never the symmetric superposition that is the system's true ground state. This and similar phenomena have been attributed \cite{harris,simonius} to a manifestation of the QZE. The idea is that merely by virtue of being in solution the molecules are constantly buffeted about and ``observed," or decohered. The time scale for this is the inverse of the energy split between the theoretical symmetric and asymmetric states of the isomers, which is expected to be extremely long (hence the decay is subject to interruptions on the time scale of collisions in solution). But as remarked in the last paragraph, this situation does not distinguish between the various characteristic times, since all are the same.

\subsection{Implications of the bound in measurements} \draftlabel{implications} \label{sec:implications}

As just shown, no quantum system, under unitary evolution alone, can become orthogonal to itself in less than $\tpass$, where $\tpass$ is, up to a trivial factor, what we have called $\tzeno$. In particular, for a given state, $\psi$, and given Hamiltonian, $H$, 
\beq
\tpass = \frac{\pi\tzeno}2
  = \frac\pi2 \frac{\hbar}{\sqrt{\langle\psi|(H-E_\psi)^2 |\psi\rangle}}
       \;. \label{passagetime}
\eeq
Moreover, we showed that for at least one system, possessing only 2-levels, the bound is actually attained.

In general measurements, however, the Fleming bound may have little to do with the time the system needs to complete its transition. Thus the Landau-Zener tunneling experiment shows transitions within the first $\mu$s, although the jump time is $\sim5\,\mu$s. In this case, since the measured $\tlife$ is $\sim70\,\mu$s, the Zeno time would presumably be the algebraic mean, $\sim20\,\mu$s. There is no doubt, from inspection of the data, that many transitions occur well before $\tpass$. How can that be? 

The answer is that proper use of the bound requires that the Hamiltonian of the measurement apparatus be included in the ``$H$" of Sec.~\ref{sec:flemingbound}. In general this can involve enormous energies, much larger than those of the system measured (were it in isolation). Thus, for the \emph{full} system $\tzeno^{\hbox{\little full}}$ may be extremely short, in particular shorter than even $\tjump$ of the isolated system. In the tunneling experiment \cite{raizen}, one has a \emph{time-dependent} Hamiltonian, reflecting the fact that controlling the value of the acceleration, $a$, as a function of time, is an important part of the successful performance of that experiment. Thus during the time that the crossed beams are turned on at their maximum $a$, the wave function of the atom in the tunneling experiment is partly in the well, partly in the barrier, partly outside. The sudden change in the confining potential means that the apparatus is interacting directly with the system, leading to a large energy spread. This remark is related to the story told to students when they first encounter barrier penetration: if you check whether the particle ``really" is in the barrier, you'd introduce enough energy to overcome that barrier. (A change in $\tzeno$ due to measurement was also seen in \cite{hinder}, but there the ``apparatus" coupling stops the decay rather than facilitates it.) 

What I now show is that for some kinds of measurement the Fleming bound provides direct physical information. Moreover, serious attention to this bound can provide an experimental test for my own theory of what takes place in a quantum measurement \cite{timebook}. 

We again consider the ``apparatus" of Sec.~\ref{sec:reconciling}. The Hamiltonian is
\draftlabel{Hwithapp2}
\beq
H=\pmatrix{ 0   & \Phi^\dagger & 0              \cr
          \Phi  & \omega       & \Theta^\dagger \cr
            0   & \Theta       & W              \cr}  \;,  \label{Hwithapp2}
\eeq
where $H$ is a $(1+N+M)\!\times(1+N+M)$-dimensional matrix; $N$ is the dimension of the diagonal matrix $\omega$, and $M$ ($\gg N$) the dimension of the diagonal matrix $W$. The states of the system are of the form $\psi^\dagger=(x^*,y^\dagger,z^\dagger)$, $x\in\compC$, $y\in\compC^N$ and $z\in\compC^M$. The physical scenario is this. The normalized state $\psi$ with $x=1$ represents an undecayed atom; call its level \#1. It is coupled, perhaps electromagnetically, to states with $y\neq0$, $z=0$, via the coupling terms $\Phi$. The ``$y$" states represent the atom in its decayed state (call it \#2) plus one or more photons. Now it may happen that the atom can continue its decay to a third level, or perhaps by varying an external field that decay can be encouraged. Let the atom in that third level, plus \emph{all} emitted photons (from both steps) correspond to the various ``$z$" levels. As in Sec.~\ref{sec:reconciling}, this second transition involves considerable decoherence and provides the irreversibility and amplification characteristic of the measurement process. Thus the way the rest of the world knows that the system has decayed from level~1 to level~2 is realized through the coupling, $\Theta$, and the states with $z\neq0$.

The important point is that \emph{for this kind of apparatus-system coupling, there is \emph{no} change in $\tzeno$}. It is still $\hbar/\sqrt{\Phi^\dagger\Phi}$. The key is that the measurement works by coupling to the decay products, \emph{not} to the original state \cite{products}, thus leaving $\tzeno$ and $\tpass$ unchanged. For such measurements, the Fleming bound does not allow the state to be completely out of its original level, nor to be completely in any other, for $t<\tpass$.

\section{Experimental discrimination among quantum measurement theories and ``special states"} \draftlabel{discrimination} \label{sec:discrimination}
\subsection{Testing the foundations} \draftlabel{testing} \label{sec:testing}

Suppose you had an apparatus of the type described in Sec.~\ref{sec:implications}, i.e., one that couples only to decay products (cf.\ Eq.\ \eqcite{Hwithapp2}). If this were a system for which $\tzeno$ is known, then one could say with confidence that unitary evolution alone cannot bring the wave function entirely to the decay states before $\tpass$. What are the implications of this according to the Copenhagen interpretation of quantum mechanics? Answer: none. You can still (for $t<\tpass$) measure the system to be in the decayed state (presumably, using \emph{this} measurement apparatus), and as usual the probability of doing so would be the absolute value squared of the amplitude in the decayed state---no need for this to be unity, just strictly positive.

By contrast, according to the explanation for the definiteness of quantum measurements that I have proposed \cite{timebook}, you would only get a definite measurement of the decay when the entire wavefunction has entered the Hilbert space of decayed states. I will not review these ideas here, and refer the reader either to the indicated book, or, for a less complete version, to \cite{def}.

This allows an explicit experimental test of my theory. A system is put in an unstable state and then shielded from the environment, except for an apparatus monitoring its state \emph{indirectly}, that is, by checking for decay products. For this system (for which I do not yet have a specific physical proposal) you would need to calculate or bound $\tpass$. I then predict \emph{no} decays before $\tpass$, whereas the Copenhagen interpretation imposes no such ban (despite some relative reduction if the system is still in the quadratic decay regime).

Although complete blocking of the environment can be difficult (cf.\ \cite{timebook}), the quest for quantum computers has in recent years developed experimental tools for just this goal. I look forward to exploring this further.

\subsection{Special states for decay} \draftlabel{special} \label{sec:special}

The motivation for this subsection is explained in \cite{timebook}. Briefly, in Sec.~\ref{sec:testing} I indicated that according to my ideas no decay could take place until $t\geq\tpass$. But what if $t=2\tpass$? Would the system \emph{then} decay, i.e., exit \emph{completely} from the state $x=1$, as my theory requires? From the Hamiltonian \eqcite{Hwithapp2} it doesn't look that way. For moderate $\Theta$ (hence $\tresp$) one gets the usual exponential decay: on a scale of $\tlife$ the wave function gradually passes out of its initial state. Since, generally, $\tlife\gg\tpass$ this implies that at $2\tpass$ most of the wave function is still in the undecayed state. My explanation for the manifest observations of decay at short times (but $>\tpass$) is that there are \emph{special} states of the environment for which the decay \emph{does} go to completion, despite the fact that for the vast majority of environmental states this does not happen. Why Nature chooses these ``special" states is discussed in \cite{timebook}. What I wish to show in the present article is a special state for decay in the model Hamiltonian \eqcite{Hwithapp2}, or in something close to it.

The physical environment is not represented in \eqcite{Hwithapp2}. The main environmental richness is in the initial state of the ambient photon field when the atom is still in its level-1, undecayed state. But this requires a cross product of available photon states with the $(1+N+M)$-dimensional Hilbert space I have heretofore considered. Instead of this, I will simplify by incorporating the field-initial-condition information in $\Theta$ itself. This quantity, in the rotating wave approximation, is of the form $\Theta= \sum_k |3\rangle\langle2| a_k^\dagger$, with $a_k^\dagger$ the photon creation operator. (Multiple photon creation is also allowed.) If the field of preexisting photons (before the decay) is well occupied, both $a_k$ and $a_k^\dagger$ can be approximated by $\sqrt{n_k}$, with $n_k$ the occupation number of the $k^{\hbox{\tiny th}}$ photon mode. This means that the features of the environment appear as particular values of the components of $\Theta$.

I have already presented something like this in \cite{scale}. I assumed that the environment fluctuates near the atom, effectively modifying $\Phi$. With a particular $\Phi(t)$ the decay \emph{is} complete by $\tpass$. However, this demonstration required beliefs about what the field could accomplish, beliefs that I did not explicitly justify.

In the present article I will show that with a purely fixed set of interactions ($\Theta$ and $\Phi$) the system will rapidly go completely over to an orthogonal state. The demonstration won't quite produce a state that makes it in $\tpass$, just $\sqrt2$ times that, but this establishes the main point.

With this in mind we break the subspace $\{(0,0,z)\}$ ($z\neq0$) into two pieces. One piece consists of a particular set of $N$ levels (one for each dimension in the space $\{(0,y^\dagger,0)\}$ with $y\neq0$). We assume that each of these has the same energy as one of the ``$y$" levels. (Recall this is the total energy, atom plus photons, so these levels correspond to the atom dropping to level-3 and emitting a photon of just the 2-3 energy difference, of which there are many.) At the same time, we assume that the occupation numbers of those levels in the ambient field are just such as to make the coupling to the ``$y$" level with energy $\omega_k$ equal to that same $\omega_k$. The coupling of the remaining degrees of freedom I call $\tTh$, and the energies $\tW$. The Hamiltonian and wave function take the form
\beq
H=\pmatrix{ 0   & \Phi^\dagger  & 0     & 0             \cr
          \Phi  & \omega        &\omega & \tTh^\dagger  \cr
            0   & \omega        &\omega & 0             \cr
            0   & \tTh          & 0     &\tW            \cr} \;,\qquad 
\psi=\pmatrix{x        \cr
              y        \cr
              \zeta    \cr
              \tilde z \cr}  \;.
\eeq
Now when most matrix elements of $\Theta$ are moderate, the passage out of the initial Hilbert subspace of undecayed atomic states is slow, on the order of $\tlife$. If it can be demonstrated that by using only the restriction of $H$ to its first $2N+1$ dimensions, one can get decay in a time on the order of $\tpass$, then the remaining couplings and levels ($\tTh$ etc.) will be negligible on that time scale. Therefore I restrict attention to the first $2N+1$ levels and study the Hamiltonian and states \draftlabel{Hspecial1}
\beq
\widehat H =
\pmatrix{ 0   & \Phi^\dagger  & 0      \cr
          \Phi  & \omega        &\omega \cr
            0   & \omega        &\omega  \cr} \;,\qquad 
\psi=\pmatrix{x        \cr
              y        \cr
              \zeta    \cr}  \;.
\label{Hspecial1}
\eeq

Two approaches will be used to analyze the dynamics. First give $\psi$ an overall dependence $\exp(-iEt)$ (with $\hbar=1$). By the same manipulations that led to \eqcite{eigvalapparatus}, $E$ is found to satisfy \draftlabel{Econdition}
\beq
E=\Phi^\dagger \frac{E-\omega}{E^2-2\omega E}\Phi 
 =\frac12\left\{ \frac1{E\tzeno^2} +\Phi^\dagger\frac1{E-2\omega}\Phi \right\}
 \label{Econdition} \;.
\eeq
As in \eqcite{gammatheta}, this becomes
\beq
E^2=\frac1{2\tzeno^2} +\frac E2\left(\Delta E -\frac i4 \Gamma \right)
 \;.
\eeq
(The denominator ``4" for $\Gamma$ arises from the $2\omega$ in \eqcite{Econdition}.) Generally both $\Delta E$ and $\Gamma$ (which is the usual decay rate) are much smaller than $1/\tzeno$, so that to a good approximation
\beq
E\approx \frac1{\sqrt2\tzeno} -\frac i{16}\Gamma
\eeq
(where $\Delta E$ is ignored). This implies that with the initial condition $x=1$ the behavior of $x$ will be $\cos(t/\sqrt2\tzeno)$ to a very good approximation. This in turn implies that $x$ will hit zero when $t=(\pi/2)\tzeno \sqrt2$. That value differs by a factor $\sqrt2$ from the optimum defined by Fleming's bound. The point though is that with a bit of manipulation of the environment the decay has been speeded up from a scale of $\tlife$ to one of $\tpass$. 

This result can also be obtained by looking at the time-dependent equations generated by the Hamiltonian of \eqcite{Hspecial1}. They are \draftlabel{schrod3}
\beq
i\dot x    =\Phi^\dagger y \;,\quad
i\dot y    =\Phi x +\omega(y+\zeta) \;,\quad
i\dot \zeta=\omega(y+\zeta)  \;. \label{schrod3}
\eeq
Add and subtract the second and third equations, integrate the equation for the difference, substitute back for $y$, and finally take the derivative with respect to $t$ to obtain \draftlabel{dynamics}
\beq
\ddot x(t) + \frac1{2\tzeno^2}x(t)
  =-\frac12 \frac{\partial}{\partial t}\int_0^t \Phi^\dagger e^{-2i\omega (t-s)} \Phi x(s)\,ds
\;. \label{dynamics}
\eeq
Define $K(u)\equiv \Phi^\dagger \exp(-i\omega u) \Phi$. This is an important kernel for studying decay properties. Thus for unobserved decay Eq.~\eqcite{schrodeq} implies $\dot x= -\int_0^t K(u)x(t-u)\,du$. Although the possibilities for $K$'s behavior are wide, for moderate times it typically drops rapidly, so that a reasonable approximation is $K(u) \approx (\Gamma/2) \delta(u)$. The normalization can be checked by plugging into the equation just written for $\dot x$. In \eqcite{dynamics} we have $K(2u)$, so that with the $\delta$-function approximation we obtain 
\beq
\ddot x(t) + \frac\Gamma8 \dot x  + \frac1{2\tzeno^2}x(t)=0    \;.
\eeq
For times less than $\tjump$ the $\delta$-function approximation is not applicable, but the Zeno time is generally much longer and is the scale now considered. With initial conditions $x(0)=1$, $\dot x=0$ (from \eqcite{schrod3}) it follows that, to lowest order in $\Gamma$,
\beq
x(t)=\cos(t/\sqrt{2}\tzeno) e^{-\Gamma t/16}  \;,
\eeq
which agrees with our previous result. An amusing perspective on the early vanishing of $x$ is as the ultimate anti-QZE \cite{reverse}.

To further confirm that the approximations work, I have included Fig.~1. This is a numerical calculation of a decay that, in the absence of apparatus-induced ``specializing" effects, would show normal exponential decay. For this calculation it is assumed that the coupling enhancement in the apparatus arising from the extra photons in the particular modes $k$ (the extra factors ``$\sqrt{n_k}$" mentioned earlier) only lasts for a period $\tpass$, after which the coupling returns to normal.

\begin{figure}[b]
\hbox{\includegraphics[width=.48\textwidth]{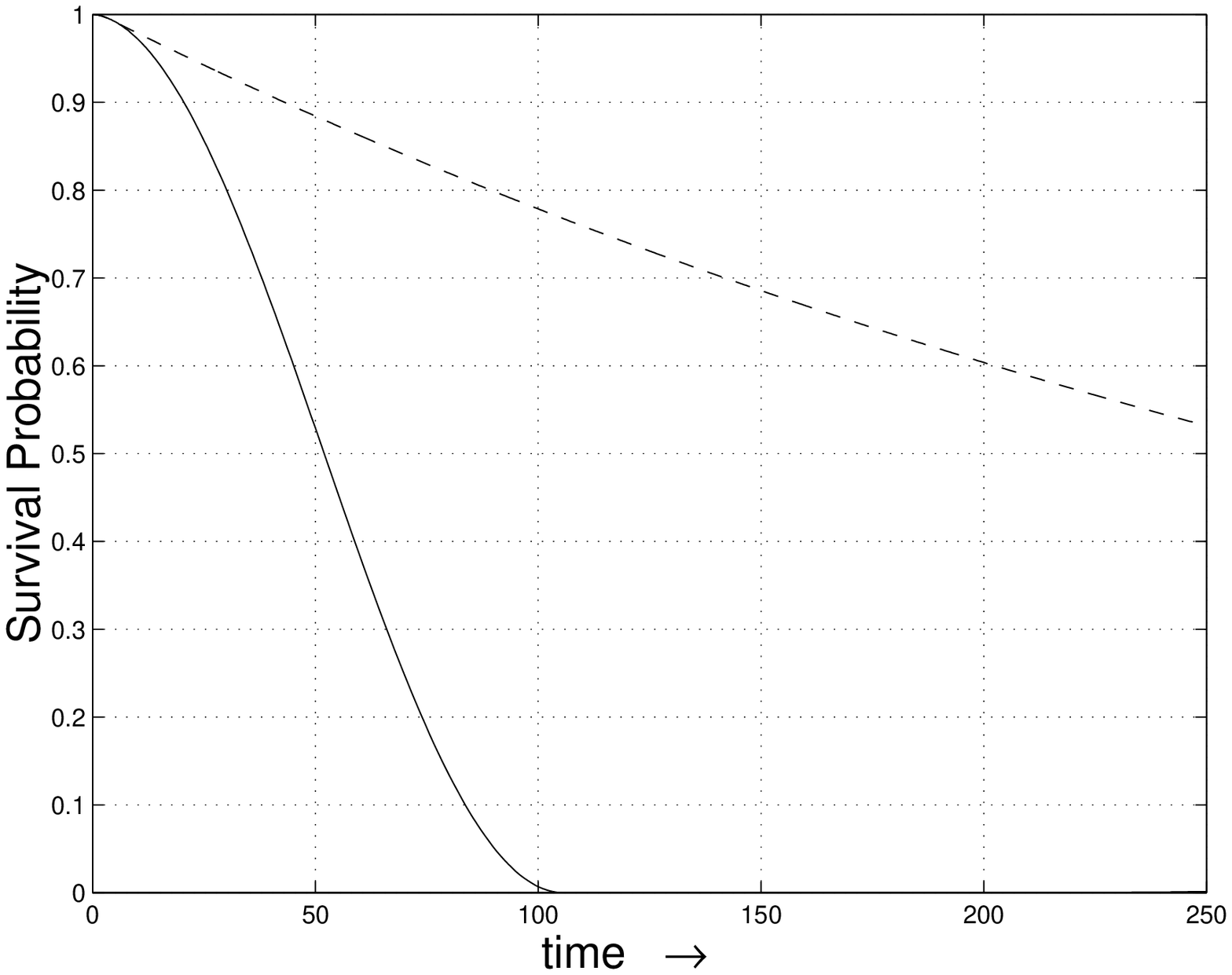}~
      \includegraphics[width=.48\textwidth]{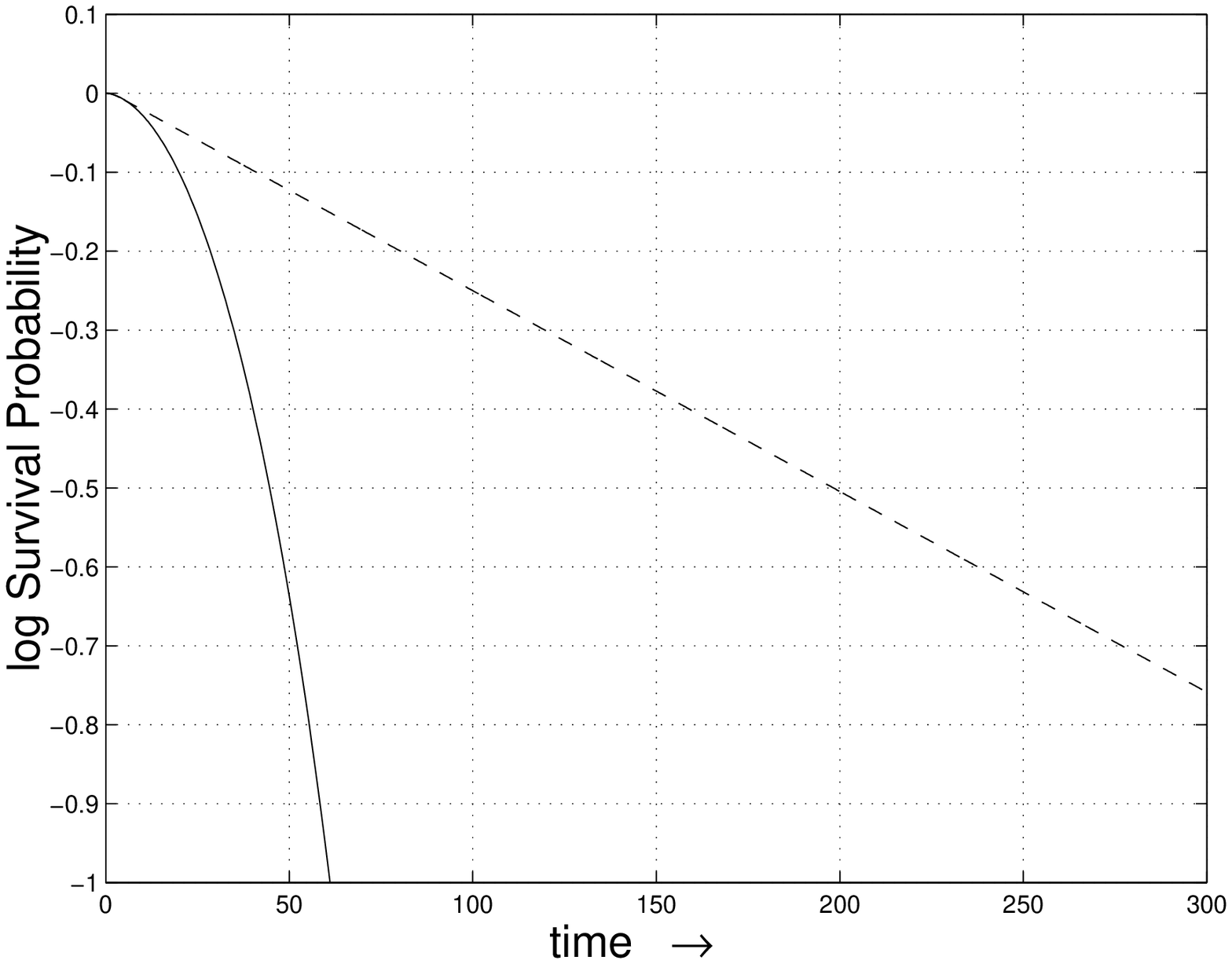}}
\caption[]{Survival probability as a function of time, linear and log plots. The solid line represents decay under the influence of the (matched photon) apparatus. The dashed line is ordinary decay, with no apparatus, and on the log plot shows appropriate linear decline. (In this system $\tzeno\approx48$ and a fit gives $\tlife \approx 393$ yielding $\tjump\approx 6$, which is too small for a deviation to be seen in this figure.) These are computer calculations of the survival probability with the Hamiltonians \eqcite{defH} and \eqcite{Hspecial1} (with transient coupling, as described in the text). From the analytic calculations, passage time should be $\sqrt2\tpass \approx 106$, and as is evident the decay in the presence of the apparatus hits zero at a time close to this. The dashed curve, representing normal decay, is far from zero at this time. In this calculation $N=101$, and continuation of the curve would eventually show quantum Poincar\'e recurrence.}
\label{fig1}
\end{figure}


This time dependence of $\Theta$ illustrates the fact that the ``specialness" of the microscopic state includes timing. The added coupling due to the ambient field is indeed ambient and once the transition is complete things return to normal. If the reader is encountering my ideas for the first time and finds the choreography excessive, please be assured that the appearance of ``unlikely" microscopic states has been addressed extensively. What is \emph{likely} or \emph{unlikely} is related to the thermodynamic arrow of time and it is by exploring related foundational questions of statistical mechanics that I am able to argue for the plausibility of these ideas. If this has piqued your interest, see \cite{timebook}.

\medskip
\remark{The states just exhibited are special states for quantum jumps. Another example, in which the environment plays an even more explicit role, is \cite{lineardecay}.}

\section{Discussion}\label{sec:discussion}\draftlabel{discussion}

Under the unitary evolution given by the formal mathematical structure of quantum mechanics, systems move gradually from state to state. For example, an unstable atom still has amplitude in its original state after many of its lifetimes. But in practice, which is to say in the lab, they go from being in one state to being in the other, seemingly instantaneously. This is the import of the term ``quantum jump." The experiments that saw single-atom transitions \cite{exp1,exp2,exp3} appear to confirm this perception. In the measurements, the system went from state to state in a time beneath the discrimination of the observers, whereas when the times spent in the unstable state were averaged, they recovered the lifetime of the atom.

The problem studied in this article is whether the ``quantum jump" is indeed instantaneous or whether it could be assigned a duration, in theory and in experiment. The longstanding problem of tunneling time, in connection with barrier penetration, sets a precedent and is a guide. If that tunneling represents a process necessary for decay, then surely the associated time is a candidate (or a lower bound) for the duration of the transition.

Two characteristic times are defined in this article, \emph{jump time} ($\tjump$) and \emph{passage time} ($\tpass$). In general they are quantitatively different and it is the richness of quantum mechanics, as well as lingering questions about its interpretation, that allow two answers to what would be a well-defined question classically.

Both times use the \emph{Zeno time}, $\tzeno$, defined in terms of the Hamiltonian of the system and its initial state as
\beq
\tzeno \equiv
   \hbar\Big/\sqrt{\langle\psi| (H-E_\psi)^2 | \psi\rangle} 
\qquad\hbox{with~} E_\psi \equiv \langle\psi|H|\psi\rangle  \;.
\eeq

The jump time, $\tjump$, takes what I consider to be a more traditional view and is defined in terms of the time scale needed to slow (\`a la the quantum Zeno effect) the decay. A ``measurement" is an idealized projection leading to
\beq
\tjump  \equiv \tzeno^2/\tlife   \;. 
\eeq
This time shows up in several contexts. It is related to tunneling time \cite{dominated}, for those transitions where a physical barrier can be identified. Its inverse is the bandwidth of the Hamiltonian, in a kind of time-energy uncertainty principle that governs the ability of the system to change state. $\tjump$ is also an indicator of the duration of the quadratic decay regime in both experiment \cite{raizen} and in numerical calculations. (An illustration of the latter is Fig.~2 of \cite{contin}.)

The passage time, $\tpass$, arises from pure unitary evolution alone, sans interpretive steps. It is based on a bound \cite{fleming} that shows that for \emph{any} $H$ and $\psi$ the system cannot evolve to a state orthogonal to $\psi$ in less time than $\tpass$, with  
\beq
\tpass=\pi \tzeno/2  \;.
\eeq
If you think of $H$ as the Hamiltonian of the system alone, than it would appear that this bound has little to do with measurements. The ``instantaneous" jump occurs outside the realm of unitary evolution (so they say) and could certainly happen faster than $\tpass$.

But I want to consider $H$ to be the Hamiltonian of \emph{the system and the measuring apparatus}. This is a view I have advocated for quite some time \cite{def1,def2,def} and is the perspective taken by the many-worlds and decoherence interpretations of quantum mechanics.

But even among those who accept this view, there is still no consensus about the implication of $\tpass$ for an actual measurement. In my theory \cite{timebook} this bound implies that the apparatus could \emph{not} detect a transition in less than $\tpass$, making this the ultimate transition time. (And in the present article an example was given of a special state that did manage the transition in close to $\tpass$.) Of course, one can still have detection in times less than the $\tpass$ you would calculate using the system Hamiltonian alone, since the full passage time of system plus apparatus, $\tzeno^{\hbox{\little full}}$, is in general much shorter than the restricted one.

Finally, there is a particular kind of detection in which the presence of the apparatus does not change the passage time. This provides the possibility of an experimental test of my measurement theory. One of my current goals is to find a practical experimental setup in which this test can be made.

\section*{Acknowledgments}
I am grateful to P. Facchi, E. Mihokova and S. Pascazio for helpful discussions. This work is supported in part by the U. S. NSF under grant PHY 97 21459.



\begin{thebibliography}{8.}
\bibitem{otherscenario} Another scenario is for the ratio \eqcite{bandratio} to grow to a distant maximum. This is the situation for the inverse QZE, mentioned above \cite{reverse}.
\bibitem{products} The term ``decay products" refers to states in Hilbert space. It is not the physical atom that is or is not a decay product---in fact, the atom in level~1, 2 or 3, appears in every state.
\end{thebibliography}
\end{document}